\documentclass{iopart}

\usepackage{graphicx}

\usepackage{iopams}

\usepackage{setstack}

\usepackage{epstopdf}
\usepackage{color}

\usepackage[T1]{fontenc}
\usepackage{textcomp}

\usepackage{mathptmx}
\usepackage{microtype}

\usepackage[scaled=.92]{helvet}

\usepackage[colorlinks=true,
		citecolor=blue,
		linkcolor=blue,
		anchorcolor=blue,
		filecolor=blue,
		menucolor=blue,
		urlcolor=blue,
		unicode
		]{hyperref}
\usepackage[all]{hypcap}

\newcommand{\mycomment}[1]{}

\newcommand{\genp}{\eta}
\newcommand{\genq}{\xi}

\newcommand{\eqref}[1]{(\ref{#1})}

\definecolor{green}{rgb}{.2,.7,0}

\begin{document}

\title{Symplectic integrators with adaptive time steps}

\author{A~S~Richardson and J~M~Finn}
\address{T-5, Applied Mathematics and Plasma Physics, Los Alamos National Laboratory, Los Alamos, NM, 87545, USA}
\ead{asrichardson@lanl.gov, finn@lanl.gov}

\date{\today}

\begin{abstract}
In recent decades, there have been many attempts to construct symplectic
integrators with variable time steps, with rather disappointing results.
In this paper we identify the causes for this lack of performance,
and find that they fall into two categories. In the first, the time step
is considered a function of time alone, $\Delta=\Delta(t)$.  In this case, backwards error analysis shows that while the algorithms remain symplectic, parametric instabilities arise because of resonance between oscillations of $\Delta(t)$ and the orbital motion.
In the second category the time step is a function
of phase space variables $\Delta=\Delta(q,p)$.  In this case, the system of equations to be solved is analyzed 
by introducing a new time variable
$\tau$ with $dt=\Delta(q,p) d\tau$.
The transformed equations are no longer in Hamiltonian form, and thus are not guaranteed to be stable even when integrated using a method which is symplectic for constant  $\Delta$.  We analyze two methods for integrating the transformed equations which do, however, preserve the structure of the original equations.  
The first is an extended phase space method, which has been successfully used in previous studies of adaptive time step symplectic integrators.
The second, novel, method is based on a non-canonical mixed-variable generating function.
Numerical trials for both of these methods show
good results, without parametric instabilities or spurious growth
or damping. It is then shown how to adapt the time step to an error estimate
found by backward error analysis, in order to optimize the time-stepping
scheme. Numerical results are obtained using this formulation and
compared with other time-stepping schemes for the extended phase space
symplectic method.
\end{abstract}
\submitto{\PPCF}

\section{Introduction}

Adaptive time integration schemes for ODEs are well established and
perform extremely well for many applications. However, for applications
involving the integration of Hamiltonian systems, there are good reasons
for using symplectic integrators\cite{0951-7715-3-2-001}. This is particularly true in applications
such as accelerators, in which very long orbits must be integrated.
Symplectic integrators are also used in particle-in-cell (PIC) codes for plasma
applications, where for economy integrators with lower order of accuracy
must be used, but it is important that the Hamiltonian phase
space structure be preserved. The same is true for molecular dynamics codes\cite{MD-1,MD-2}, which are of use for dense plasmas as well as other applications. Similarly, for tracing magnetic field lines (e.g., in tokamaks or reversed field pinches) or doing RF (radio frequency) ray tracing, symplectic integrators can be very useful.  To our knowledge, the symplectic integrators in use for plasma and particle beam applications all have uniform time-stepping. However, for these applications there are situations in which variable or adaptive time stepping could increase efficiency greatly. For example, in accelerators, particles can experience fields that vary over a wide range of length scales. The same is true for for PIC and MD codes, for example for particles moving in and out of shocks or magnetic reconnection layers, or undergoing close collisions.

There have been studies of symplectic
integrators with variable time steps, but the early results were not promising \cite{SkeelGear,skeel_1993,CalvoSanz-Serna93,lee1997variable,Wright1998421,Skeel1998758,gdc_bf00048485}.
Among these studies, there have been two types of variation of time
steps. In the first set of studies, the time step varies with time explicitly, and
in some cases varies from $\Delta t=\Delta_{1}$ to $\Delta t=\Delta_{2}$
on alternate blocks of time steps. In these studies problems were observed to arise. In the second set of studies, the time step
was chosen with $\Delta t=\Delta(q,p)$, where $q,p$ are the dynamical variables. For any adaptive scheme for an autonomous system of equations, the time step is of this form. For the $\Delta(q,p)$ case, the equations are no longer in canonical Hamiltonian form (but may be Hamiltonian in non-canonical variables; see below) and indeed unfavorable results are found.
The disappointing results for both of these cases have contributed to the general impression that if you need an adaptive time step integrator, you are better served using a high order non-symplectic integration method.%
\footnote[1]{An exception to this is the extended phase space approach of Hairer\cite{Hairer1997219} and Reich\cite{reich1999}, which we discuss in later sections.
Some authors\cite{1995ApJ...443L..93H} also suggest that methods which are reversible under some symmetry $R$  ($RT(h)R=T(h)^{-1}$, where the one-time-step map is $T(h)$ and $R$ is an involution, $R^2=$ the identity) should be considered in place of symplectic integrators. We do not consider reversible methods here because their advantages are limited to systems where the original flow also has this symmetry property.
}

In this paper we take a new look at symplectic integrators with adaptive
time stepping. Our focus is not on developing the most efficient
or accurate symplectic integrators, but to understand and solve the
problems that have been encountered in using symplectic integrators
with variable time steps, and to outline a method of obtaining an optimal time-step adaptation scheme.

In Sec.~\ref{sec:h_t} we start by considering time steps depending explicitly
on time alone as $\Delta t=\Delta(t) = \Delta_0 (1+\epsilon \cos \omega t)$. 
In Sec.~\ref{sec:h_t_analytic} we consider several examples of
first and second order symplectic integrators applied to the harmonic
oscillator, and apply {\em modified equation analysis} or {\em backward error analysis}\cite{Hirt1968339,Auerbach1991189,Hairer1997219,reich1999}. This allows us to find 
the modified Hamiltonian system which the integration scheme, including the effect of numerical errors, actually solves.
From this analysis we find that resonances between $\omega$ and the oscillator frequency $\Omega_0$ drive parametric instabilities. 
The resonances have $\omega=m\Omega_{0}$, the lowest
order resonance having $m=2$. For first order schemes the resonance
width scales as $\epsilon\Delta_{0}$, while second order schemes have width
$\propto\epsilon\Delta_{0}^{2}$. These parametric instabilities explain
the problematic results obtained in the papers dealing with $\Delta=\Delta(t)$\cite{skeel_1993,lee1997variable,Wright1998421,Skeel1998758}. 
In the context of the harmonic oscillator, these problems were
identified as arising from parametric instabilities by Pich\'e\cite{piche98}.

We continue in Sec.~\ref{sec:h_t_numerical} by numerically investigating time step variations
$\Delta(t)=\Delta_{0}(1+\epsilon \cos \omega t)$, by integrating the harmonic oscillator with a symplectic method.  We find parametric instabilities with resonance widths as predicted by the modified equation analysis.  We show results of numerical integrations of a cubic oscillator with the nonlinear potential $V(q)=q^2/2+ q^{3}/3$, in which the frequency varies with respect to the action variable,
$\Omega=\Omega(J)$. The parametric instabilities seen globally for
the harmonic oscillator show up as nonlinear resonances or islands,
where $\omega=m\Omega(J)$, with $m$ depending on the integration method and the Hamiltonian being integrated.   We observed $m=1$ islands for the cubic oscillator, integrated with the Crank-Nicolson (CN) method.  The calculations of this section illustrate the point that, when considering $\Delta=\Delta(t)$,  problems arise because the equations being solved become unstable (but still Hamiltonian), and not because the integration scheme fails to be symplectic.

In Sec.~\ref{sec:d_qp} we study time step variations $\Delta = \Delta(q,p)$.
With $\Delta(q,p) \propto 1/\rho(q,p)$, the straightforward substitution
$\rho(q,p)dt=d\tau$ leads to equations (having $\tau$ as the independent
variable) that are not Hamiltonian equations in canonical form. Symplectic integrators  therefore should not be expected to give stable results in this case, and again it has been observed that they do not. 
In Sec.~\ref{sec:hairer} we review one approach to solving this problem, by extending the phase
space $(q,p)\rightarrow(q,q_{0},p,p_{0})$, where $q_0=t$ and $p_0$ is its conjugate momentum. After this transformation
to an extended phase space, it is easily shown that the resulting
equations are Hamiltonian in canonical form and can be integrated by any fixed time step ($\Delta \tau=h=const.$) 
symplectic integrator in this extended phase space. This is a well-known
method, and has been used to obtain symplectic integration schemes
with variable time steps by Hairer\cite{Hairer1997219}, and also Reich\cite{reich1999}.  Hairer described this approach
as a {}``meta-algorithm'' because any symplectic integrator can
be used in the extended phase space. 
We then present numerical results in Sec.~\ref{sec:ext_num} for the cubic oscillator (with two different symplectic integrators) which show that, 
even with a step size variation that appears to have an $m=1$ resonance, evidence of non-symplectic behavior occurs and no resonant islands form.

In Sec.~\ref{sec:noncan} we introduce a second method for dealing with the problems that arise when $\Delta=\Delta(q,p)$. The point of introducing a second method is to illustrate that the main problem with $\Delta(q,p)$ is that the equations are no longer Hamiltonian, and that there are potentially many ways to address this problem.  The extended phase space method ``fixes'' the equations by embedding them in a higher dimensional Hamiltonian system.  This alternative approach does not extend the dimension of the phase space, but rather recognizes the fact that, with $\tau$ as
the independent variable and $\rho(q,p) dt=d\tau$, the equations are Hamiltonian equations
in \emph{non-canonical variables}\cite{Morrison2006593}. (This is true only for one degree of freedom, i.~e.~ 2D phase space.)
That is, the equations can
be expressed in terms of a non-canonical bracket\cite{Morrison2006593}. We write
the non-canonical variables as $(x,y)$ with the numerical scheme giving the mapping
$(x,y)=(x(\tau),y(\tau)) \to (X,Y)=(x(\tau+h),y(\tau+h))$. Equivalent to the non-canonical
bracket, the fundamental two-form is $\rho(x,y)dx\wedge dy$ rather
than the canonical form $dq\wedge dp$. We describe a method for constructing
a non-canonical generating function $F(x,Y)$ which gives a 
map $T_{F}(h)$ which is a first order accurate integrator that preserves the two-form, and discuss
its relation with transforming to canonical variables (guaranteed to be
possible by the Darboux theorem\cite{Morrison2006593}.) We then find a complementary non-canonical
generating function $G(X,y)$ which leads to another non-canonical
map $T_{G}(h)$, which is also first order accurate and preserves the two-form. 
Maps preserving the two-form $\rho(x,y) dx \wedge dy$ are called {\em Poisson integrators} \cite{Zhong1988134,GE:1990fk,Channell199180,Karasozen20041225} and conserve the measure $\int \rho \, dx \, dy$ rather than the area $\int dq \, dp$.
We show that $T_{F}(h)^{-1}=T_{G}(-h)$ and vice-versa,
so that the composition $
T(h)\doteq T_{G}(h/2)\circ T_{F}(h/2)$
is time symmetric ($T(h)^{-1}=T(h)$) and therefore second order accurate.
Finally, we  show numerical results in Sec.~\ref{sec:NSLF_num} using the composed map $T(h)$, applied to the cubic oscillator.  As with the extended phase space method, this method is stable, with errors in the energy remaining bounded.

In Sec.~\ref{sec:h_qp_error_est} we study the choice of time step variation $\rho(q,p)$. We begin by using backward error analysis for a
slightly different purpose than in Sec.~\ref{sec:h_t}, namely to estimate the
form of the error of an integrator, and then devise a scheme to minimize
the total error. This scheme takes the form of an \emph{equidistribution principle}, 
of the form $\rho_{opt}(q,p)dt=d\tau$, with uniform steps in $\tau$.
We then compare the error in numerical integration using $\rho_{opt}$ to the error obtained for fixed time steps, to the error obtained for time steps designed to give equal arc length per step, and to time-stepping using a weighted average of $\rho_{opt}$ and $\rho=1$. We show numerical results showing that, among these options for time step variation, $\rho_{opt}$ does indeed minimize the error.

In Sec.~\ref{sec:summary} we summarize our work. 
In \ref{sec:pert} we consider the relationship between the cases with $\Delta(t)$ and $\Delta(q,p)$.  We show that, while a simple argument motivates the use of $\Delta(t)$, this is only valid for short times.  We explain how this is analogous to the problems which arise in perturbation theory, where a perturbation causes a frequency shift, but at lowest order it appears a secularity or an instability.
In \ref{sec:app_a} we review non-canonical variables, brackets, and forms, material necessary for Sec.~\ref{sec:noncan}.

\section{Integration with variable time step $\Delta(t)$}
\label{sec:h_t}

In this section we will investigate symplectic integrators with a variable time step $\Delta=\Delta(t)$, in order to understand better the results that have been obtained in the literature.  Using  modified equation analysis, we will see that the problems that have been reported are best understood as a parametric instability due to resonance between the system dynamics and the time dependence of $\Delta(t)$.

\subsection{Modified equation analysis and parametric instabilities}
\label{sec:h_t_analytic}
We first look at the Harmonic oscillator $H=p^{2}/2+\Omega_{0}^{2}q^{2}/2$.
With the change of variables $p\rightarrow p/\sqrt{\Omega_{0}}$ and $q\rightarrow\sqrt{\Omega_{0}}q$
we obtain $H=\Omega_0(p^{2}+q^{2})/2$. The leapfrog (LF) method, which is first order accurate when $q$ and $p$ are not taken to be staggered), with $\Delta=\Delta(t)$ gives
\begin{eqnarray}
q_{i+1}&=q_{i}+\Delta(t_{i})\Omega_{0}p_{i},\\
p_{i+1}&=p_{i}-\Delta(t_{i})\Omega_{0}q_{i+1}.
\end{eqnarray}
Modified equation analysis begins by expanding $q_{i+1}=q_i +\Delta(t_i)\dot{q}_i +\Delta(t_i)^2\ddot{q}_i /2+\cdots$ and similarly for $p_{i+1}$. We obtain
\begin{eqnarray}
\dot{q}&=\frac{\Delta(t)\Omega_{0}^{2}}{2}q+\Omega_{0}p+O(\Delta^{2}),\\
\dot{p}&=-\Omega_{0}q-\frac{\Delta(t)\Omega_{0}^{2}}{2}p+O(\Delta^{2}).
\end{eqnarray}
(Terms proportional to $\dot{\Delta}$ appear at the next order for first order leapfrog. For other integration schemes, the $\dot \Delta$ terms will similarly appear at one order higher than the order of the method.) This system of equations arises from the Hamiltonian
\begin{equation}
H'=\frac{\Omega_{0}}{2}\left(p^{2}+q^{2}\right)+\frac{\Delta(t)\Omega_{0}^{2}}{2}qp.\end{equation}
As expected, the modified equations are in Hamiltonian form \cite{Yoshida1990262,calvo1994modified,Auerbach1991189,Benettin:1994,hairer1994backward,reich1999}. With the action-angle transformation $q=\sqrt{2J} \cos \phi,\,\, p=-\sqrt{2J} \sin \phi$
and setting $\Delta(t)=\Delta_{0}\left(1+\epsilon \cos \omega t\right)$
we obtain
\begin{equation}
 H'=\Omega_{0}J\left(1-\frac{\Omega_{0}\Delta_{0}}{2} \sin 2\phi\right) 
-\frac{\Omega_{0}^{2}\Delta_{0}\epsilon}{4}J\left[ \sin (2\phi-\omega t)+ \sin (2\phi+\omega t)\right].
\end{equation}
Keeping only the resonant $O(\Delta_{0})$ term $\propto \sin (2\phi-\omega t)$,
we obtain
\begin{equation}
H'\approx\Omega_{0}J-\Omega_{0}J\frac{\Omega_{0}\Delta_{0}\epsilon}{4} \sin (2\phi-\omega t).
\end{equation}
The canonical transformation $\psi=\phi-\omega t/2,\,\, K'=H'-\omega J/2$
($J$ unchanged) transforms to a rotating frame in phase space and gives
\begin{equation}
K'=\left(\Omega_{0}-\frac{\omega}{2}\right)J-\frac{\Omega_{0}^{2}\Delta_{0}\epsilon}{4}J \sin 2\psi.\end{equation}
We conclude from this backward error analysis that the harmonic oscillator
with time step $\Delta(t)=\Delta_{0}\left(1+\epsilon \cos \omega t\right)$
is unstable if
\begin{equation}\label{eq:lf_stab_1}
\left|\Omega_{0}-\frac{\omega}{2}\right|<\frac{\Omega_{0}^2\Delta_{0}\epsilon}{4},
\end{equation}
since $K'$ is hyperbolic in this case.
This instability is a parametric instability associated with the resonance
$\omega=2\Omega_{0}$.  Figure~\ref{fig:lf_stab} shows this predicted region of instability as a function of $\omega$ and $\epsilon$. Numerical results for comparison are shown in the next section.
Carrying out this analysis to one higher order in $\Delta_0$ gives an $O(\Delta_0^2)$ correction to the $\phi$- and time-independent terms in $K'$,
which gives rise to a positive $O(\Delta_0^2)$ shift in the resonant frequency independent of $\epsilon$, discussed in the next section.

Another integration scheme is symmetrized leapfrog, which is second order accurate, and takes the form
\begin{eqnarray}
q_{*}&=&q_{i}+\frac{\Delta(t_{i})\Omega_{0}}{2}p_{i}, \nonumber\\
p_{i+1}&=&p_{i}-\Delta(t_{i})\Omega_{0}q_{*},\\
q_{i+1}&=&q_{*}+\frac{\Delta(t_{i})\Omega_{0}}{2}p_{i+1}.\nonumber
\end{eqnarray}
This leads to a modified equation\begin{equation}
\dot{q}=\Omega_{0}p-\frac{\Delta(t)^{2}\Omega_{0}^{3}}{12}p,\,\,\,\,\,\dot{p}=-\Omega_{0}q-\frac{\Delta(t)^{2}\Omega_{0}^{3}}{6}q,\end{equation}
with modified Hamiltonian\begin{equation}
H'=\Omega_{0}\left(1-\frac{\Delta(t)^{2}\Omega_{0}^{2}}{12}\right)\frac{p^{2}}{2}
+\Omega_{0}\left(1+\frac{\Delta(t)^{2}\Omega_{0}^{2}}{6}\right)\frac{q^{2}}{2}.
\end{equation}
A similar analysis to that above for first order LF leads to
\begin{equation}
K'=\left(\Omega_{0}+\frac{\Omega_0^3 \Delta_0^2}{24} -\frac{\omega}{2}\right)J+\frac{\Omega_{0}^{3}\Delta_{0}^{2}\epsilon}{8}J \cos 2\psi.
\end{equation}
Note that for this method, the frequency shift enters at the same order as the resonant term.
We conclude that parametric instability occurs for
\begin{equation}
\left|\Omega_{0} +\frac{\Omega_0^3 \Delta_0^2}{24}  -\frac{\omega}{2}\right|<\frac{\Omega_{0}^{3}\Delta_{0}^{2}\epsilon}{8}.
\end{equation}
This instability is centered around $\omega=m(\Omega_{0} +\Omega_0^3\Delta_0^2/24)$, again with
$m=2$, but in a narrower range $\sim\epsilon\Delta_{0}^{2}$. (Resonances with other values of $m$ enter at higher order in $\Delta_0$ and $\epsilon$.)  
Also note that, because the frequency shift enters at the same order as the resonant term, the shift is comparable to the resonance width for $\epsilon\sim 1$.

As another second order accurate example, we consider Crank-Nicolson
(CN),
\begin{equation}
q_{i+1}=q_{i}+h(t_{i})\Omega_{0}\frac{p_{i}+p_{i+1}}{2},\,\,\,\,\, p_{i+1}=p_{i}-h(t_{i})\Omega_{0}\frac{q_{i}+q_{i+1}}{2},
\end{equation}
for which the modified equation analysis gives
\begin{equation}
\dot{q}=\left(1-\frac{\Delta(t)^{2}\Omega_0^2}{12}\right)\Omega_{0}p,\,\,\,\,\,\dot{p}=-\left(1-\frac{\Delta(t)^{2}\Omega_0^2}{12}\right)\Omega_{0}q
\end{equation}
or\begin{equation}
H'=\left(1-\frac{\Delta(t)^{2}\Omega_0^2}{12}\right)\Omega_{0}\left(\frac{p^{2}+q^{2}}{2}\right)=\Omega(t)J.\end{equation}
Here, $H'$ is independent of $\phi$ and the action $J=(p^{2}+q^{2})/2$ is an invariant to order $\Delta_0^2$, and the errors
are all in the phase $\phi$: the frequency $\Omega(t)=1-\Delta(t)^{2}\Omega_0^2/12$ is downshifted for this case. We conclude that parametric instabilities
cannot occur to this order for CN. It is easily seen that CN preserves
the action $J$ {\em exactly} for the harmonic oscillator; this shows that we have $H'=\Omega(t)(p^{2}+q^{2})/2$
{\em to all orders}. Further, the explicit time dependence disappears if
we introduce $\tau$ such that $\Omega(t)dt=d\tau$.

\subsection{Numerical results with $\Delta=\Delta(t)$}
\label{sec:h_t_numerical}
\begin{figure}[htbp]
\begin{center}
\includegraphics[width=8cm]{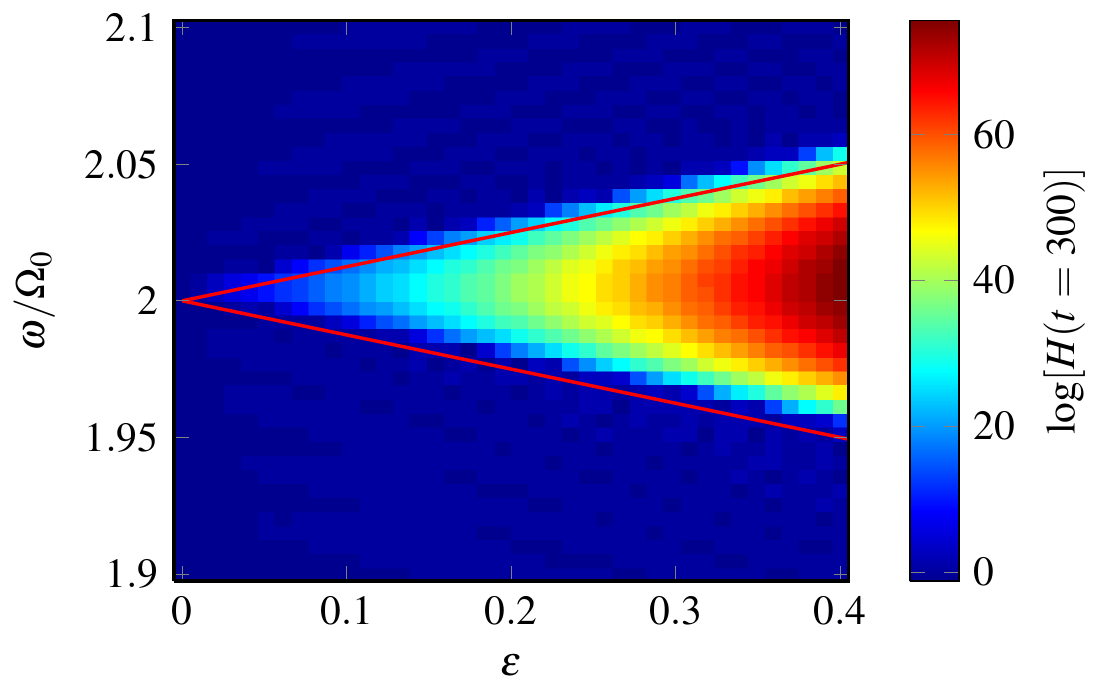}
\caption{\label{fig:lf_stab}
Numerical results showing a region of instability for a range of values of the parameters $\epsilon$ and $\omega$, for first order LF. The red lines show the predicted stability limits given in Eq.~\eqref{eq:lf_stab_1}.  The colors show the logarithm of the Hamiltonian function $H$ evaluated on the orbit after integrating the equations to $t=300$.  This value should be approximately constant along the orbits, but it becomes large on orbits that spiral out to large values of $q$ and $p$. Parameters used: $\Omega_0 = 5$, $\Delta_0=0.05$. Initial conditions: $q=0.5$, $p=0$.
The small upward shift in $\omega$ comes in at second order in $\Delta_0$.
}
\end{center}
\end{figure}

\begin{figure}[htbp]
\begin{center}
\includegraphics[width=7cm]{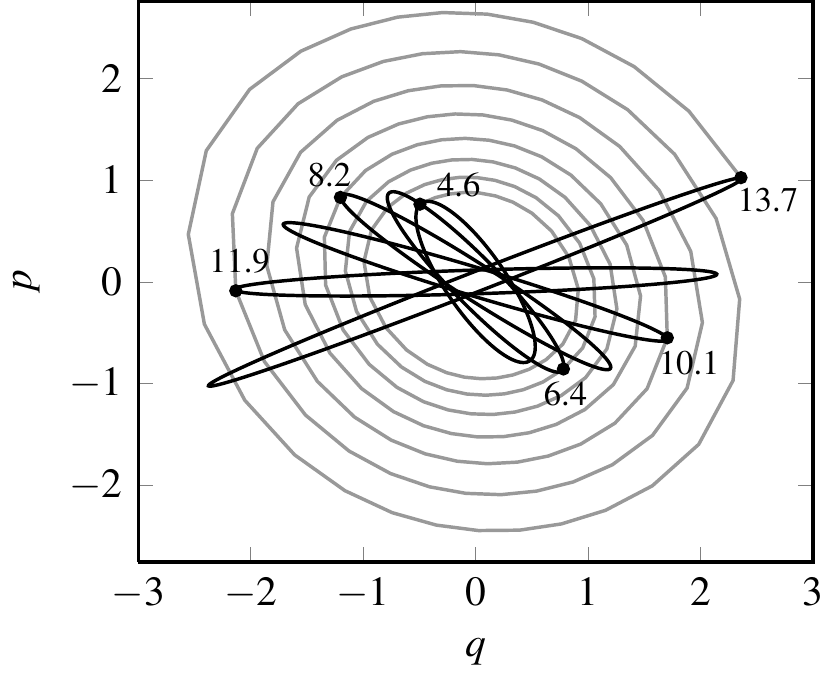}
\caption{\label{fig:ellipse}
Numerical results illustrating that, while the first order LF method is area preserving, variable time steps can cause a parametric instability. 
The black ellipses are generated by taking a set of points, and propagating them forward in time using first order LF.  The black dots show one of these points, which is plotted for various times (as labeled).  The grey curve shows the coordinates $(q(t), p(t))$ of this point, for each of the computed time steps.
Parameters: $\epsilon=0.4$, $\omega=10$, $\Omega_0=5$, $\Delta_0=0.05$.
}
\end{center}
\end{figure}

The parametric resonances described for the harmonic oscillator in the previous section were studied numerically, and the results are shown in Figs.~\ref{fig:lf_stab} and \ref{fig:ellipse}.  Figure \ref{fig:lf_stab} shows the numerical results, using the first order LF method, again with time steps given by $\Delta(t) = \Delta_0 (1+\epsilon \cos \omega t)$.  The red lines show the boundaries of the unstable region in $\omega$, $\epsilon$ parameter space, as predicted in Eq.~\eqref{eq:lf_stab_1}. The colors show the logarithm of the value of the Hamiltonian at the end ($t=300$) of a numerically computed orbit.  For unstable parameter values, the orbit spirals outward, and the final value of the Hamiltonian becomes exponentially large.  The numerically computed region of instability agrees well with the prediction of Eq.~\eqref{eq:lf_stab_1}, and also shows the $O(\Delta_0^2)$ offset in frequency. 
In Fig.~\ref{fig:ellipse} we examine how, for parameters where the integration is unstable, a circle of initial conditions is stretched out into a long thin ellipse. Since the LF method is still symplectic, the area of the ellipses is preserved. 

As described above, first order LF integration of the harmonic oscillator with variable time step $\Delta (t) = \Delta_0 (1 + \epsilon \cos \omega t)$ was found to be unstable near the resonance $\omega = 2 \Omega_0$. For a nonlinear oscillator, the oscillation frequency $\Omega_0$ depends on the amplitude.  We thus  expect that the numerical integration of a nonlinear oscillator would exhibit resonant islands, rather than exhibiting global parametric instability.

For example, consider a Hamiltonian with a cubic potential
\begin{eqnarray}\label{eq:cub_ham}
H(q,p) =  \frac{q^2 + p^2}{2} + \frac{q^3}{3}.
\end{eqnarray}
This Hamiltonian has an O-point at $(q,p)=(0,0)$ and an \mbox{X-point} at $(-1,0)$.  Inside the separatrix of the X-point, this system exhibits nonlinear oscillations.  For this system and first order LF, the time step variation $\Delta(t) = \Delta_0 (1+ \epsilon \cos \omega t)$ introduces an $m=1$ resonance ($\omega = \Omega_0$), which is readily understood by backward error analysis. Specifically, this analysis gives a term in the modified Hamiltonian which is proportional to $\sin\phi \cos\omega t$, which leads to the $m=1$ resonance.

\begin{figure}[htbp]
\begin{center}
\includegraphics[width=7cm]{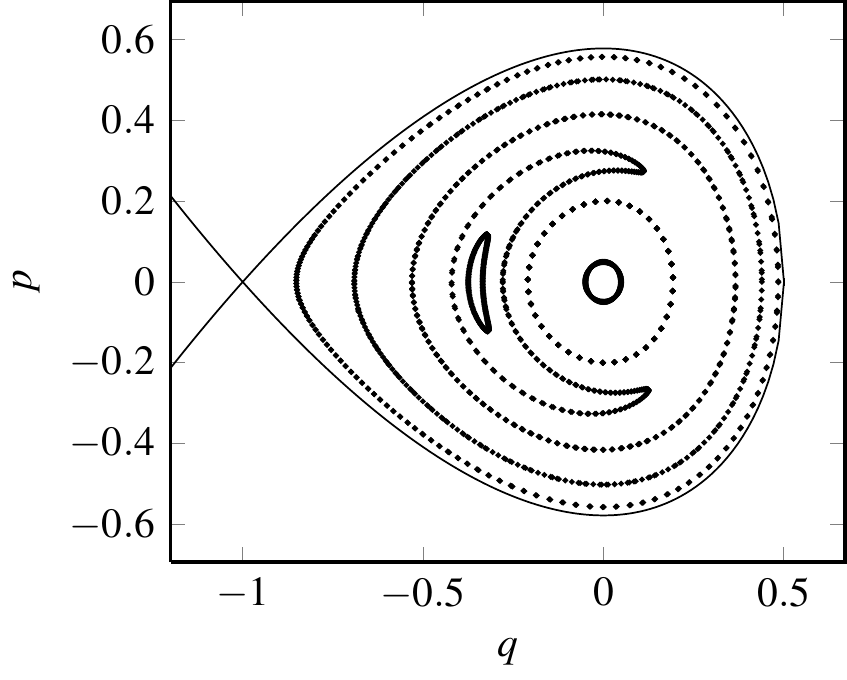}
\caption{\label{fig:cn_cub_island}
Surface of section plot for the cubic Hamiltonian of Eq.~\eqref{eq:cub_ham}, with variable time step $\Delta (t) = \Delta_0 (1 + \epsilon \cos \omega t)$.  The resonance with the time-step variation causes an $m=1$ island to open.  Parameters used: $\Delta_0 = 0.3$, $\epsilon=0.6$, $\omega=0.95$.  The dots are points in the surface of section (computed using CN), and the thin line is the separatrix of the X-point at $(-1, 0)$. 
}
\end{center}
\end{figure}

Figure~\ref{fig:cn_cub_island} shows the results of integrating this system with the CN method, and the same time steps variation.  The frequency $\omega$, $0 < \omega < 1$, was chosen so that it would resonate with an orbit inside the separatrix. (The frequency varies with amplitude from the O-point from a maximum $\Omega=1$ at the O-point to $\Omega=0$ at the separatrix.) The dots in Fig.~\ref{fig:cn_cub_island} show the Poincar\'e surface of section for a time--$T$ map, with $T=2\pi/\omega$.  The elliptic point (O-point) of a resonant $m=1$ island is clearly seen near $(q,p)=(-0.35, 0)$.
We see that indeed the time dependent step size variations lead to an unphysical parametric resonance  for the Hamiltonian system in Eq.~\eqref{eq:cub_ham}.

\section{Integration with variable time step $\Delta(q,p)$}
\label{sec:d_qp}

We now turn from considering the time step variation $\Delta(t)$ to variations of the form $\Delta(q,p)$.  This form of variation is perhaps more natural, since it arises when trying to optimize the time step to reduce errors in the integration method applied to an autonomous system. (We will return to this issue in Sec.~\ref{sec:h_qp_error_est}.)  In this section, we first discuss the problems that arise with this variation, and then we describe two methods by which these problems can be resolved.  

Previous attempts to apply this type of time step variation to symplectic integrators have suffered from one main, although sometimes unrecognized, problem: the equations are no longer Hamiltonian.
As has been pointed out\cite{Hairer1997219,reich1999}, integration of Hamiltonian's equations
\begin{equation}
\frac{dz_{i}}{dt}=\epsilon_{ij}\frac{\partial H(\mathbf{z},t)}{\partial z_{j}},\label{eq:canonical}
\end{equation}
with variable time steps satisfying $\rho(q,p) dt=d\tau$ is equivalent to changing the time variable $t$ to $\tau$, and integrating the following equations with fixed size steps in $\tau$; $\Delta \tau = h$:
\begin{eqnarray}\label{eq:non-canonical}
\frac{dz_{i}}{d\tau}&=\frac{1}{\rho(\mathbf{z})}\epsilon_{ij}\frac{\partial H(\mathbf{z},t)}{\partial z_{j}},\\ 
\frac{dt}{d\tau} &=  \frac{1}{\rho({\bf z})}, \nonumber
\end{eqnarray}
where  $\epsilon_{12}=-\epsilon_{21}=1,\,\,\,\epsilon_{11}=\epsilon_{22}=0$ (summation assumed).
Here, $\mathbf{z}=(q,p)$. As they stand, these equations are not in canonical Hamiltonian form.  Because of this, the application of  a symplectic integrator should not be expected to give results that are symplectic, since Eq.~\eqref{eq:non-canonical} is not a canonical Hamiltonian flow.

Once this problem has been identified, there are various approaches that can be used to find an integrator which preserves the Hamiltonian nature of Eq.~\eqref{eq:canonical}, even with time steps given by $\rho(q,p)dt = d\tau$.
In the remainder of this section, we review and discuss two methods for integrating Eqs.~\eqref{eq:non-canonical} in such a way that the symplectic nature of the original equations [Eq.~\eqref{eq:canonical}] is preserved.  The first method is to embed these equations into a larger phase space, considering $t$ to be another coordinate. 
This extended phase space approach goes back to the work of Sundman \cite{sundman}, and has more recently been applied to numerical methods, both non-symplectic \cite{springerlink:10.1007/BF01650285} and symplectic \cite{Hairer1997219,reich1999}.
In the extended phase space, the new equations can again be written in canonical Hamiltonian form, and thus integrated with any (fixed $\tau$ step) symplectic integrator.  
This extended phase space method is not the only way to integrate Eqs.~\eqref{eq:non-canonical} symplectically, however.
A second method which we consider in this section (which only applies to one degree of freedom problems) is to recognize the the equations can be written in terms of a non-canonical bracket.  Any method which preserves this bracket will then have all the nice properties of a symplectic integrator for a canonical system.  We present a novel generating function approach for constructing such a method.

\begin{figure}[htbp]
\begin{center}
\includegraphics[width=7cm]{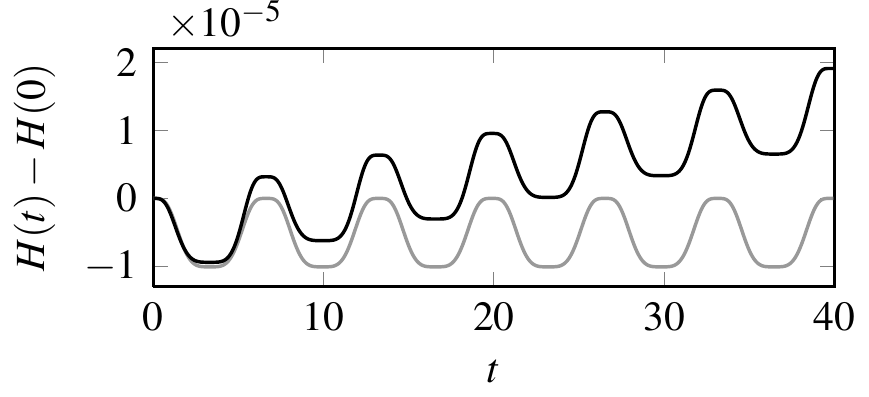}
\caption{\label{fig:compare_cn_nonuniform}
CN integration of the cubic Hamiltonian [Eqs.~\eqref{eq:cub_ham} and \eqref{eq:canonical}] is shown in gray, and the time transformed version of the cubic Hamiltonian [Eqs.~\eqref{eq:cub_ham} and \eqref{eq:non-canonical}, with $\rho$ as given in Eq.~\eqref{eq:varstep_ex}] is shown in black.  Parameters: $h=0.05$, $\Delta_0=1.05134$, $a_1=0.5$, $a_2=0.25$.  Initial conditions: $(q_0,p_0)=(0.3,0)$. 
}
\end{center}
\end{figure}

For the numerical results presented in this section, we will consider the Hamiltonian with a cubic potential given in Eq.~\eqref{eq:cub_ham}, integrated with the two different methods.  The variable time steps $\Delta = \Delta(q,p)$ are chosen so they might appear to have an $m=1$ resonance, since that is what is expected to give islands similar to the previous results of Fig.~\ref{fig:cn_cub_island}.  Let%

\begin{eqnarray}
\Delta(q,p) = \Delta_0\left(1 + a_1 q + a_2 p\right).
\end{eqnarray}
For the simulations reported in this paper, $a_1=0.5$ and $a_2 = 0.25$.%
\footnote[1]{We take $a_2\neq 0$ in order to make the equations asymmetric in $p$.  Otherwise, the symmetry of CN makes the results appear symplectic, when in general they would not be.
} 
For $\rho(q,p)\Delta(q,p)=h$, i.e.~$\Delta t=\Delta(q,p)$ and $\Delta \tau=h$, we have
\begin{eqnarray}\label{eq:varstep_ex}
\rho(	q,p) = \frac{h}{\Delta(q,p)} = \frac{h}{\Delta_0\left(1+a_1 q + a_2 p\right)}.
\end{eqnarray}
The step size $\Delta \tau =h$ was held fixed, and $\Delta_0$ was adjusted so that after a given number $N$ periods of the orbit, $t(N)$ equals $\tau(N)$.  This is done so that the differences between the original and extended phase space results are due to the step size variation, rather than a change in the average step size.
Integrating the original Hamilton equations \eqref{eq:canonical} with fixed step size CN  (i.e., with $a_1 = a_2 = 0$) gives a result with bounded errors in energy [see Fig.~\ref{fig:compare_cn_nonuniform}], as is expected for symplectic integrators. However, integrating Eqs.~\eqref{eq:canonical} using CN with step size given by Eq.~\eqref{eq:varstep_ex} gives errors in energy with linear growth, as also seen in Fig.~\ref{fig:compare_cn_nonuniform}.  This lack of boundedness of the energy is to be expected, since varying the step size in this way is equivalent to integrating Eqs.~\eqref{eq:non-canonical} with fixed step size, and \eqref{eq:non-canonical} is not a Hamiltonian system.  Thus there is no reason that using a symplectic method such as CN would give results with bounded errors in energy.
(Note that the Hamiltonian is just a useful diagnostic; the underlying problem is that phase space area is not conserved.)

\subsection{Extended phase space method}
\label{sec:hairer}

By a well-known procedure \cite{Hairer1997219,reich1999,sundman,springerlink:10.1007/BF01650285}, we introduce $q_{0}=t$ and its canonically
conjugate momentum $p_{0}$ and consider the Hamiltonian in extended
phase space:
\begin{equation}
K(q,p,q_{0},p_{0})=\frac{1}{\rho(\mathbf{z})}\left(H(q,p,q_{0})+p_{0}\right),
\label{eq:Extended-Hamiltonian}
\end{equation}
again with $\mathbf{z}=(q,p)$. The canonical equations of motion are 
\begin{eqnarray}
\frac{dq}{d\tau}&=\frac{1}{\rho(\mathbf{z})}\frac{\partial H(q,p,q_0)}{\partial p}+\left(H+p_{0}\right)\frac{\partial}{\partial p}\left(\frac{1}{\rho}\right), \nonumber \\
\frac{dp}{d\tau}&=-\frac{1}{\rho(\mathbf{z})}\frac{\partial H(q,p,q_0)}{\partial q}-\left(H+p_{0}\right)\frac{\partial}{\partial q}\left(\frac{1}{\rho}\right),\label{eq:Extended-canonical-Ham} \\
\frac{dq_{0}}{d\tau}&=\frac{\partial K}{\partial p_{0}}=\frac{1}{\rho(\mathbf{z})},\nonumber \\
\frac{dp_{0}}{d\tau}&=-\frac{\partial K}{\partial q_{0}}=-\frac{1}{\rho(\mathbf{z})}\frac{\partial H}{\partial q_{0}}.\nonumber
\end{eqnarray}
The last of these equations implies that if we choose $p_{0}(\tau=0)=-H(q(0),p(0),0)$, then ${dp_0}/{d\tau}=-{dH}/{d\tau}$ and
$H+p_{0}$ remains zero. In this case, Eqs.~(\ref{eq:Extended-canonical-Ham})
are identical to Eqs.~(\ref{eq:non-canonical}). 
So, with this choice of initial condition for $p_0$, the actual orbit is followed and $K=0$.  Note that, for this value of $p_0$, the other contours of $K=\rm{const.}$ are not orbits of the original system.

With the equations of motion in the form (\ref{eq:Extended-canonical-Ham}),
we can then use a canonical symplectic integrator such as Modified LF (ML) (a semi-implicit, symmetrized version of LF; see \cite{finn:054503,fichtl2011}) or CN with fixed time step $\Delta\tau=h$. This
is the method used in Refs.~\cite{Hairer1997219,reich1999} to integrate Hamiltonian
equations of motion with variable time steps for the special case
$\partial H/\partial t=0$, for which $H$ and $p_{0}$ are separately
constants of motion.
It should be noted that in a symplectic integration
of Eqs.~(\ref{eq:Extended-canonical-Ham}), $H+p_{0}$ will be only
approximately invariant, so that the results of numerically integrating
Eqs.~(\ref{eq:Extended-canonical-Ham}) will have small errors relative
to the numerical results of integrating Eqs.~(\ref{eq:non-canonical}).

Note that if $H$ is independent of $t$, then one needs only to integrate the equations for $q$ and $p$, since $p_0$ is exactly constant, and $q_0=t$ can be found by post-processing.  So, even though this method ``extends'' phase space, the dimension of the system effectively remains unchanged. Therefore, there is no possibility of parametric instability if the time step is small enough.

\mycomment{
{\color{red}This explains why no resonances. For extended phase space in 1 deg freedom: backward error analysis gives $K=K(q,p,p_0)$, i.e. no $q_0$ dependence, so $p_0$ is just a constant, a parameter. So it's still 1 deg freed and $K(q,p) = const$ labels the orbits, so no resonances are possible.
}
}

\subsection{Numerical results: extended phase space method}
\label{sec:ext_num}

\begin{figure}[htbp]
\begin{center}
\includegraphics[width=7cm]{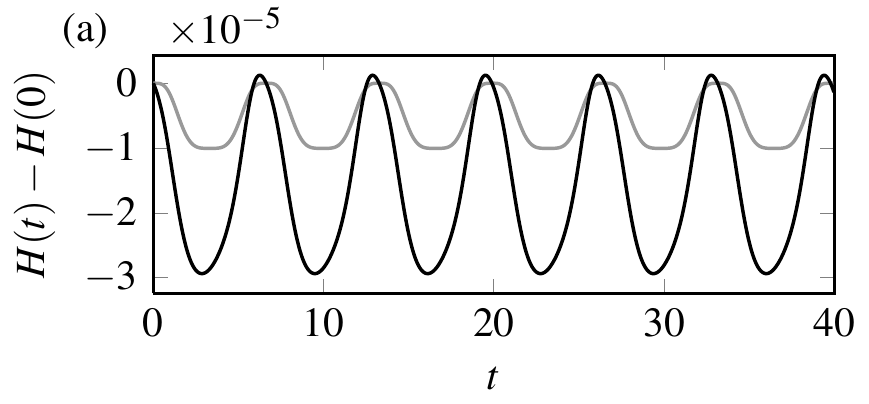}
\includegraphics[width=7cm]{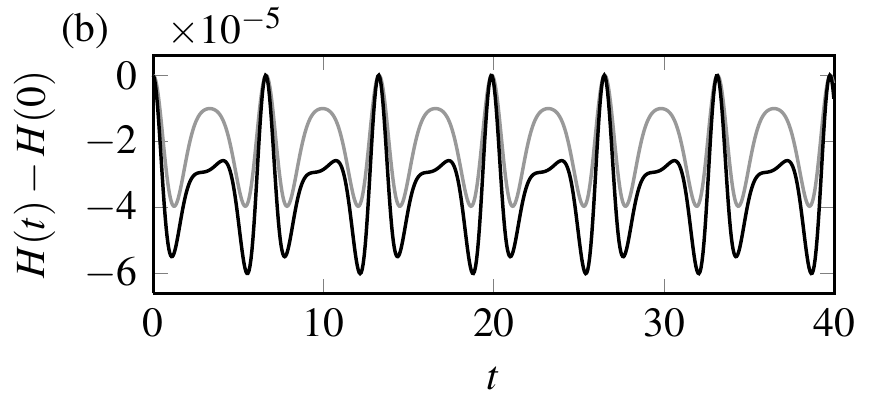}
\caption{\label{fig:compare_cn_ext}
Extended Phase Space Method.
Comparison of (a) CN and (b) ML.
Shown in black is the error in energy $H(q(t),p(t))$ for the two methods.  For comparison, in gray is shown the error in energy for the same method, but applied to the original uniform time step discretization of the Hamiltonian equations.
Parameters: $h=0.05$, $\Delta_0=1.05117$, $(q_0,p_0)=(0.3,0)$, $a_1 = 0.5$, $a_2 = 0.25$.
}
\end{center}
\end{figure}

In this section we report numerical results using the extended phase space method [Eq.~\eqref{eq:Extended-canonical-Ham}] to integrate the equations from the cubic Hamiltonian   [Eq.~\eqref{eq:cub_ham}], using $\rho(q,p)$ as given in Eq.~\eqref{eq:varstep_ex}.
In Fig.~\ref{fig:compare_cn_ext} the energy errors are compared for (a) CN integration and (b) ML integration.  In both cases, the black curve gives the extended phase space results, while the gray curve gives the result obtained by applying the same method with fixed step size to the original Hamiltonian equations.  
As for Fig.~\ref{fig:compare_cn_nonuniform}, the $\tau$ step $h$ was chosen to make $\tau(N)$ equal to $t(N)$ after $N$ periods of the orbit.
We conclude that there is no secular change in the value of the Hamiltonian $H(q(t),p(t))$. 
The time step in Eq.~\label{eq:varstep_ex} was chosen because it appears at first to have to potential for an $m=1$ resonance; in Sec.~\ref{sec:h_qp_error_est} we will describe a method to minimize the integration errors.

\begin{figure}[htbp]
\begin{center}
\includegraphics[width=7cm]{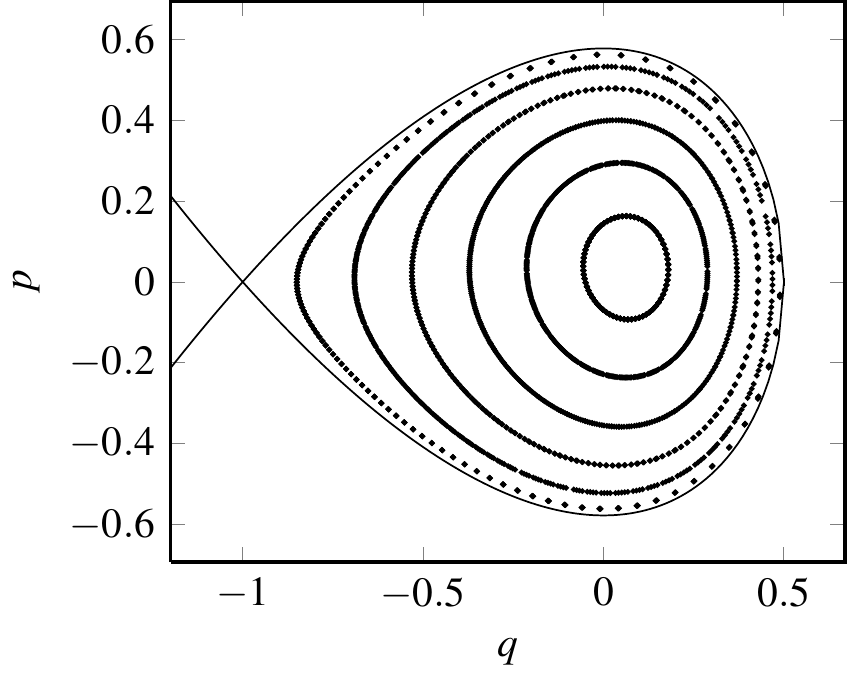}
\caption{\label{fig:ext_cub_island}
Surface of section plot, strobed at an arbitrary period $T$, for the extended phase space method (computed using CN).
Parameters: $a_1 = 0.5$, $a_2 = 0.25$, $\Delta_0=0.3$.
Expansion of the scale for results from longer runs show no islands anywhere inside the separatrix for any period $T$.
}
\end{center}
\end{figure}

In Fig.~\ref{fig:ext_cub_island} is shown a surface of section plot for the extended phase space CN integration.  In contrast to Fig.~\ref{fig:cn_cub_island}, there is no resonant island caused by the time step variation. This is consistent with the comment at the end of Sec.~\ref{sec:hairer}. 

As a final check of the extended phase space method, a comparison was made between integrating Eqs.~\eqref{eq:non-canonical} where the step size is determined by the position in phase space as $\Delta(q,p) \propto 1/\rho(q,p)$, and integrating Eq.~\eqref{eq:canonical} where the step size is a function of time as determined by a reference orbit $\Delta(t) \propto 1/\rho(q_{ref}(t), p_{ref}(t))$.  The reference orbit was obtained by solving the extended phase space equations [Eqs.~\eqref{eq:non-canonical}], for some initial conditions $(q_0, p_0)$ for which the period is $T(q_0,p_0)$.  Time-$T$ surface of section plots for several cases give a results qualitatively similar to Fig.~\ref{fig:cn_cub_island}, with the time dependent step size $\Delta(t)=\Delta(q_{ref}(t), p_{ref}(t))$ giving a resonant island centered at the location of the reference orbit. This behavior is similar to the secular and unstable cases of perturbation theories discussed in \ref{sec:pert}.

\subsection{Non-canonical symmetrized leapfrog}
\label{sec:noncan}
In this section we discuss a second method which can be used to integrate Eqs.~\eqref{eq:non-canonical}, which works for one degree of freedom systems (2D phase space). In this section, we change the notation for a point in phase space from $(q,p)$ to ${\bf z}=(x,y)$.  This is done to emphasize the fact that Eqs.~\eqref{eq:non-canonical} are not in canonical Hamiltonian form.

Equations~\eqref{eq:non-canonical} are of the form $\mathbf{u}=\mathbf{m}/\rho$, where
$\mathbf{u}$ is the velocity, with $\nabla\cdot\mathbf{m}=\nabla\cdot\rho\mathbf{u}=0$.
The measure $\int_{V(t)}\rho dV$ is preserved over any phase space
volume $V(t)$ advected with the flow. Equivalently, for the time--$t$
map $\mathbf{z}(0)\rightarrow\mathbf{z}(t)$ we have%
\footnote[1]{For tracing magnetic field lines by $dx_{i}/d\zeta=B_{i}/B_{\zeta}$
($i=1,2$) the same considerations lead to conservation of flux  $\int_{A}B_{\zeta}dA$ at
two ends of a flux tube, and ${\rm det}\left(\partial x_{i}(\zeta)/\partial x_{j}(0)\right)=B_{\zeta}(\mathbf{x}(0))/B_{\zeta}(\mathbf{x}(\zeta))$.
}%
\begin{equation}
{\rm det}\left(\frac{\partial z_{i}(t)}{\partial z_{j}(0)}\right)=\frac{\rho(\mathbf{z}(0))}{\rho(\mathbf{z}(t))}.\label{eq:determinant-condition}
\end{equation}
This measure preservation suggests that a measure preserving integrator applied to this system would have the same nice properties of a symplectic integrator applied to a canonical Hamiltonian system.

The equations of motion in the form in Eq.~(\ref{eq:non-canonical})
are not canonical Hamiltonian equations. However, in one degree of freedom, Eq.~(\ref{eq:non-canonical}) is a Hamiltonian system in non-canonical variables and
can be written in the form\begin{equation}
\frac{dz_{i}}{d\tau}=[z_{i},H]_{z},\,\,\,\,\,{\rm where}\,\,\,\,\,[f,g]_{z}=\frac{1}{\rho(\mathbf{z})}\epsilon_{ij}\frac{\partial f}{\partial z_{i}}\frac{\partial g}{\partial z_{j}}\end{equation}
is a non-canonical bracket. That is, it is antisymmetric and direct calculation shows that it satisfies the Jacobi
identity $[f,[g,h]_{z}]_{z}+[g,[h,f]_{z}]_{z}+[h,[f,g]_{z}]_{z}=0$.

A method of constructing a symplectic non-canonical integrator in
terms of generating functions\cite{Islas2001116} starts with the condition
that must hold for the time--$t$ map, which is that it must preserve
the two-form \begin{equation}\label{eq:preserve}
\rho(x,y)dx\wedge dy=\rho(X,Y)dX\wedge dY,
\end{equation}
See  \ref{sec:app_a} for more details regarding this condition.
First, we define 
\begin{equation}
\genp (x,y)=\int^{y}\rho(x,y')dy'\label{eq:canonical-p}
\end{equation}
and
\begin{equation}
\genq (x,y)=\int^{x}\rho(x',y)dx'.\label{eq:canonical-Q}
\end{equation}
With these definitions, we can write
\begin{eqnarray}
\rho(x,y)dx\wedge dy&=dx\wedge d\genp(x,y), \\
\rho(X,Y)dX\wedge dY&=d\genq(X,Y)\wedge dY
\end{eqnarray}
and therefore Eq.~\eqref{eq:preserve} becomes
\begin{equation}\label{eq:preserve2}
dx\wedge d\genp(x,y)=d\genq(X,Y)\wedge dY.
\end{equation}
We can consider $(X,Y)$ to be functions of $(x,y)$, and then define $p=\genp(x,y)$ and $Q=\genq(X,Y)$. These definitions, together with Eq.~\eqref{eq:preserve2}, imply that the one-form
\begin{equation}
\omega=p dx+Q dY
\end{equation}
is exact, i.e.~$d\omega=0$.  This in turn implies
\begin{equation}
\omega=dF(x,Y)=\frac{\partial F(x,Y)}{\partial x}dx+\frac{\partial F(x,Y)}{\partial Y}dY,\end{equation}
at least locally. We are led to the relations
\begin{eqnarray}
\genp(x,y)&=p=\frac{\partial F(x,Y)}{\partial x},\label{eq:F2-eq-1}\\
\genq(X,Y)&=Q=\frac{\partial F(x,Y)}{\partial Y}.\label{eq:F2-eq-2}
\end{eqnarray}
Given the generating function $F(x,Y)$, we
solve Eq.~(\ref{eq:F2-eq-1}) for $Y(x,y)$ and then substitute into
Eq.~(\ref{eq:F2-eq-2}) to obtain $X(x,y)$. 

In essence, what is going on is this: we are transforming to canonical
variables $(x,p)$ and $(Q,Y)$, where $p=\genp(x,y)$ and $Q=\genq(X,Y)$. Then we are performing
a canonical transformation $(x,p)\rightarrow(Q,Y)$ in terms
of a generating function $F(x,Y)$, with $Q=\partial F/\partial Y$ and $p=\partial F/\partial x$, and expressing the results in terms
of $(x,y)$, and $(X,Y)$. 

The identity transformation has $X=x$ and $Y=y$ so that Eqs.~(\ref{eq:F2-eq-1}),
(\ref{eq:F2-eq-2}) take the form
\begin{eqnarray}
\genp(x,y)&=\frac{\partial F_{0}(x,y)}{\partial x},\label{eq:F2-eq-1-identity} \\
\genq(x,y)&=\frac{\partial F_{0}(x,y)}{\partial y}.\label{eq:F2-eq-2-identity}
\end{eqnarray}
These relations ensure that $\partial \genp(x,y)/\partial y=\partial \genq(x,y)/\partial x=\partial^{2}F_{0}/\partial x\partial y$;
by Eqs.~(\ref{eq:canonical-p}), (\ref{eq:canonical-Q}), these both
equal $\rho(x,y)$, which is nonzero. This nondegeneracy is necessary
for the inversions in Eqs.~(\ref{eq:F2-eq-1}), (\ref{eq:F2-eq-2})
to work.

We now wish to construct an $O(h)$ approximation to $F(x,Y)$ in order to find a time--$h$ map ($h=\Delta\tau)$
for the system given by Eq.~(\ref{eq:non-canonical}).  Let us try
the generating function
\begin{equation}
F(x,Y)=F_{0}(x,Y)+hH(x,Y).
\end{equation}
From Eqs.~(\ref{eq:F2-eq-1}) and (\ref{eq:F2-eq-2}) we find
\begin{eqnarray}
\genp(x,y)&=\frac{\partial F_{0}(x,Y)}{\partial x}-hv(x,Y),\label{eq:F2-for-leapfrog1}\\
\genq(X,Y)&=\frac{\partial F_{0}(x,Y)}{\partial Y}+hu(x,Y),\label{eq:F2-for-leapfrog2}
\end{eqnarray}
where $u(x,y)=\partial H(x,y)/\partial y$ and $v(x,y)=-\partial H(x,y)/\partial x$.
Substituting Eqs.~(\ref{eq:F2-eq-1-identity}) and (\ref{eq:F2-eq-2-identity}) into these equations
gives
\begin{equation}
\label{eq:ML-form-1}
\boxed{
\genp(x,y)=\genp(x,Y)-hv(x,Y),\quad
\genq(X,Y)=\genq(x,Y)+hu(x,Y).
}
\end{equation}
This is the form of our integrator: we integrate $\rho$ to give
$\genp(x,y)$ and $\genq(x,y)$ as in Eqs.~(\ref{eq:canonical-p}) and (\ref{eq:canonical-Q}),
iterate the first of (\ref{eq:ML-form-1}) to solve for $Y$ and,
after substituting for $Y$, iterate the second of (\ref{eq:ML-form-1})
to solve for $X$. We do not need the explicit form for $F_{0}$ or the Hamiltonian
$H$.

In order to see that Eqs.~\eqref{eq:ML-form-1} are a first order accurate integrator for the system in Eq.~\eqref{eq:non-canonical}, we expand to first order in $h$, and find
\begin{eqnarray}
0=(Y-y)\frac{\partial \genp(x,y)}{\partial y}-hv(x,y)+O(h^{2})\\
(X-x)\frac{\partial \genq(x,y)}{\partial x}=hu(x,y)+O(h^{2}).
\end{eqnarray}
With Eqs.~(\ref{eq:canonical-p}), (\ref{eq:canonical-Q}), this leads
to
\begin{equation}
X=x+h\frac{u(x,y)}{\rho(x,y)}+O(h^{2}),\quad Y=y+h\frac{v(x,y)}{\rho(x,y)}+O(h^{2}).
\label{eq:first-order-form}
\end{equation}
This implies that Eqs.~(\ref{eq:ML-form-1}) 
do constitute a first order accurate integrator for Eqs.~(\ref{eq:non-canonical}).
It is clear that the scheme in Eqs.~(\ref{eq:ML-form-1}) is the non-canonical form of the ML integrator.\cite{fichtl2011,finn:054503}
Note, however, that the system in Eqs.~\eqref{eq:ML-form-1} is implicit in both $X$ and
$Y$.

For the complementary generating function $G(X,y)$ we follow the
same procedure, defining  $\genq(x,y)$ and $\genp(x,y)$ exactly as in Eqs.~\eqref{eq:canonical-p} and \eqref{eq:canonical-Q}. We can now write Eq.~\eqref{eq:preserve} as
\begin{equation}
d\genq(x,y)\wedge dy=dX\wedge d\genp(X,Y),
\end{equation}
which, together with the definitions $P=\genp(X,Y)$ and $q=\genq(x,y)$,
implies that \begin{equation}
\omega'=-q dy-P dX\end{equation}
is exact ($d\omega'=0$). Therefore $\omega'=dG(X,y)=(\partial G(X,y)/\partial X)dX+(\partial G(X,y)/\partial y)dy$
and
\begin{equation}
\genq(x,y)=-\frac{\partial G(X,y)}{\partial y}, \quad
\genp(X,Y)=-\frac{\partial G(X,y)}{\partial X}.
\end{equation}
The identity is given by $G_{0}(X,y)$, where
\begin{equation}
\genq(x,y)=-\frac{\partial G_{0}(x,y)}{\partial y},\quad
\genp(x,y)=-\frac{\partial G_{0}(x,y)}{\partial x},
\end{equation}
with $-\partial^{2}G_{0}/\partial x\partial y=\rho(x,y)$. We
try
$G(X,y)=G_{0}(X,y)+hH(X,y),$
which leads to
\begin{equation}
\label{eq:ML-form-2}
\boxed{
\genq(x,y)=\genq(X,y)-hu(X,y),\quad
\genp(X,Y)=\genp(X,y)+hv(X,y).
}
\end{equation}
As in the case with Eqs.~\eqref{eq:ML-form-1}, this is the form we use as an integration scheme, and it is not necessary to obtain
$G_{0}$ explicitly. Finally, a first order expansion in $h$ again
gives Eqs.~(\ref{eq:first-order-form}), showing that Eqs.~\eqref{eq:ML-form-2} are also a first order accurate integrator for Eqs.~(\ref{eq:non-canonical}).

The first order maps $(x,y)\rightarrow(X,Y)$ defined by Eqs.~(\ref{eq:ML-form-1}) and Eqs.~(\ref{eq:ML-form-2})
involve integration of a one-degree of freedom Hamiltonian system with no explicit time dependence,
and thus do not exhibit a parametric instability for small enough time steps. 

We have shown that both schemes in Eq.~(\ref{eq:ML-form-1}) and
in Eq.~(\ref{eq:ML-form-2}) are first order accurate integrators
for Eq.~(\ref{eq:non-canonical}). What remains is to show that if
they are applied sequentially one obtains second order accuracy. The
first point is that if the maps based on $F(x,Y)$ and $G(X,y)$ are
written as $T_{F}(h)$ and $T_{G}(h)$, respectively, one can show
that\begin{equation}
T_{F}^{-1}(h)=T_{G}(-h)\end{equation}
and of course $T_{G}^{-1}(h)=T_{F}(-h)$. This follows from inspection
of Eqs.~(\ref{eq:ML-form-1}) and (\ref{eq:ML-form-2}).

The composed map is $T(h)=T_{G}(h/2)\circ T_{F}(h/2)$ and we find
\begin{equation}
 T(h)^{-1}=T_{F}(h/2)^{-1}\circ T_{G}(h/2)^{-1}
=T_{G}(-h/2)\circ T_{F}(-h/2)=T(-h).
\end{equation}
Thus, $T(h)$ is reversible and
is therefore second order accurate.\footnote[1]{%
Straightforward backward error analysis shows that, if a single-step map $T(h)$ is accurate to first order and time-symmetric [i.e.~satisfies $T(h)^{-1}=T(-h)$], then it must be accurate to at least second order.
} We call $T(h)=T_G(h/2) \circ T_F(h/2)$ {\em non-canonical symmetrized leapfrog} (NSL).

The methods outlined in this section preserve the non-canonical bracket,
as does the actual time evolution of the non-canonical system. Integrators
with this property are often called \emph{Poisson integrators}\cite{Zhong1988134,GE:1990fk,Channell199180,Karasozen20041225}.

It is interesting to note in this context that (fixed step size) CN can be written as $T_{CN}(h)= T_f (h/2) \circ T_b(h/2)$, where $T_f$ is forward (explicit) Euler and $T_b$ is backward (implicit) Euler. Further, the implicit midpoint method $x'=x+(h/2)u(x)+(h/2)u(x')$ can be written as $T_{mid}(h) =  T_b(h/2) \circ T_f(h/2)$.  This shows that $T_{mid}=T_f(h/2)^{-1} \circ T_{CN}(h) \circ T_f(h/2)$, so that the implicit midpoint method is related to CN by a non-canonical change of variables. This implies that the implicit midpoint method is equivalent to a symplectic method under a (non-symplectic) change of variables, so it is a Poisson integrator. Therefore it can be used to integrate a canonical Hamiltonian system with all the advantages of a symplectic integrator.

\subsection{Numerical results: non-canonical symmetrized leapfrog}
\label{sec:NSLF_num}

\begin{figure}[htbp]
\begin{center}
\includegraphics[width=7cm]{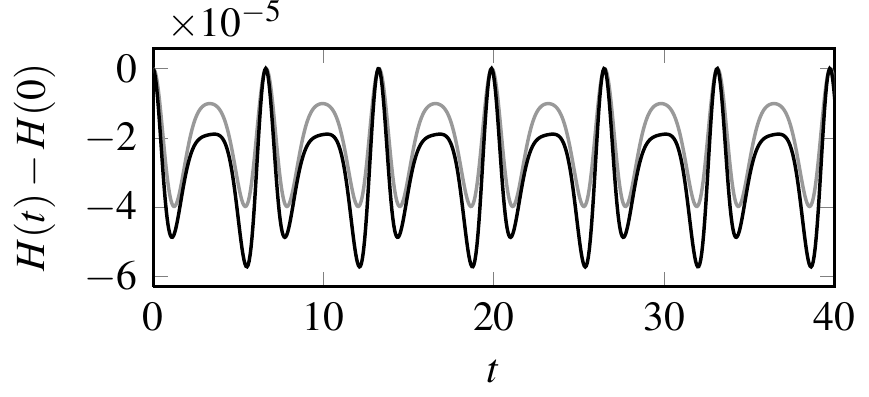}
\caption{\label{fig:compare_poisson}
The NSL method applied to the cubic oscillator, with time step variations as given in Eq.~\eqref{eq:varstep_ex}.
Shown in black is the error in energy.  For comparison, in gray is shown the error in energy for fixed step size symmetrized LF.
Parameters: $h_0=0.05$, $\Delta_0=1.0524$, $(q_0,p_0)=(0.3,0)$, $a_1=0.5$, $a_2=0.25$.
}
\end{center}
\end{figure}

In this section we report numerical results where the non-canonical symmetrized leapfrog (NSL) integration scheme is applied to the cubic Hamiltonian [Eq.~\eqref{eq:cub_ham}], with $\rho(x,y)=h/\Delta_0 (1+a_1 x+a_2 y)$, as given in Eq.~\eqref{eq:varstep_ex}.
Integrating $\rho(x,y)$ yields  $\genp(x,y) = h\ln (1+a_1 x+a_2 y)/\Delta_0 a_2$ and $\genq(x,y)=h \ln (1+a_1 x+a_2 y)/\Delta_0 a_1$, and thus Eq.~\eqref{eq:ML-form-1} with this Hamiltonian and time step variation becomes 
\begin{eqnarray}
\ln(1+ a_1 x + a_2 y') &= \ln(1+a_1 x + a_2 y)  +\Delta_0 a_2 v(x)  , \\
\ln(1 + a_1 x' + a_2 y') &= \ln(1+a_1 x + a_2 y')  + \Delta_0 a_1 u(y), \nonumber
\end{eqnarray}
where $u=\partial H/\partial y=y$ and $v  =-\partial H/\partial x=-(x+x^2)$. The first of these can be solved for $y'$ and then the second for $x'$, giving a non-canonical leapfrog integrator.  The second non-canonical generating function method, given in Eq.~\eqref{eq:ML-form-2}, can similarly be written in terms of logarithms. As described in the last section, the composition of these two methods results in a symmetrized integrator, which can be written as:
\begin{eqnarray}
\label{eq:non-can_lf}
y^* &= \left[ (1+a_1 x + a_2 y) e^{\Delta_0 a_2 v(x)/2} - a_1 x -1\right]/a_2, \\
x' &= \left[ (1+a_1 x + a_2 y^*) e^{\Delta_0 a_1 u(y^*)} - a_2 y^* -1\right]/a_1,\nonumber\\
y' &= \left[ (1+a_1 x'+ a_2 y^*) e^{\Delta_0 a_2 v(x')/2} - a_1 x' -1\right]/a_2, \nonumber
\end{eqnarray}
Figure~\ref{fig:compare_poisson} shows numerical results from integrating the cubic oscillator using the NSL method of Eqs.~\eqref{eq:non-can_lf} (shown in black) and  fixed step size symmetrized LF (shown in gray). As for the extended phase space method, we find that the NSL method with this $\rho(x,y)$ gives results with bounded errors in the energy. Again, this choice of $\rho(x,y)$ was chosen because it suggests $m=1$ resonance and leads to tractible equations; in Sec.~\ref{sec:num_opt} we show results with an optimized $\rho(x,y)$ that minimizes the time-stepping error. The surface of section for this method looks qualitatively like Fig.~\ref{fig:ext_cub_island}, again with no resonant islands.

\section{Using error estimates to find $\Delta(q,p)$}
\label{sec:h_qp_error_est}

We next turn our attention to the choice of step size variation $\Delta=\Delta(q,p)$,
based on an error estimator. Basing the choice of $\Delta$ on an error estimate will give step size variations which are optimized to reduce errors, in a manner to be defined.

\subsection{Error estimates and minimization}

As an example of an error estimate we consider CN
integration of $d\mathbf{z}/dt=\mathbf{u}(\mathbf{z})$, where again $\mathbf{z}=(x,y)$. Namely, we define
a map $\mathbf{z}'=T_{h}(\mathbf{z})$ given by
\begin{equation}
\mathbf{z}'=\mathbf{z}+\Delta\mathbf{u}\left((\mathbf{z}+\mathbf{z}')/2\right),
\end{equation}
where $\mathbf{z}=\mathbf{z}(t)$ and $\mathbf{z}'=\mathbf{z}(t+\Delta)$,
where $\Delta=\Delta t$ can vary  as a function of $\bf z$. If $\mathbf{u}$ comes
from a Hamiltonian system, CN is symplectic, but that property is
not important for these considerations. We can thus let $\Delta(t)=\Delta(q_{ref}(t),p_{ref}(t))$, where $q_{ref}$ and $p_{ref}$ are the exact orbits.%
\footnote[1]{
Also, since we are considering $\Delta=\Delta(t)$, we do not need the terms proportional to $H+p_0$ in the extended phase space method, Eq.~\eqref{eq:Extended-canonical-Ham}.  Regardless of the form of $\Delta$, this term is of higher order than the error we are estimating in this calculation.
}%
As mentioned in Sec.~\ref{sec:h_t_analytic}, there will be terms proportional to $d\Delta/dt$, but they will appear at the next order in $\Delta$.

The error estimate is found by expanding $\mathbf{z}(t+\Delta)$ in $\Delta$, and performing backward error analysis to obtain a local error estimate. This is a very different objective than that in the backwards error analysis in Sec.~\ref{sec:h_t}.  In index notation we obtain
\begin{eqnarray}
\frac{dz_{i}}{dt}+\frac{\Delta}{2}\frac{d^{2}z_{i}}{dt^{2}}+\frac{\Delta^{2}}{6}\frac{d^{3}z_{i}}{dt^{3}}&=u_{i}\left(\mathbf{z}+\frac{\Delta}{2}\frac{d\mathbf{z}}{dt}+\frac{\Delta^{2}}{4}\frac{d^{2}\mathbf{z}}{dt^{2}}\right) \\
&=u_{i}(\mathbf{z})+\frac{\Delta}{2}\dot{z}_{j}u_{i,j}+\frac{\Delta^{2}}{4}\ddot{z}_{j}u_{i,j}+\frac{\Delta^{2}}{8}\dot{z}_{j}\dot{z}_{k}u_{i,jk}. \nonumber
\end{eqnarray}
This leads to
\begin{equation}
\frac{dz_{i}}{dt}=u_{i}+\Delta^{2}\left(\frac{1}{12}u_{k}u_{j,k}u_{i,j}-\frac{1}{24}u_{j}u_{k}u_{i,jk}\right),
\end{equation}
which can be written in vector notation as
\begin{equation}
\frac{d\mathbf{z}}{dt}=\mathbf{u}+\Delta^{2}\left(\frac{1}{12}(\mathbf{u}\cdot\nabla\mathbf{u})\cdot\nabla\mathbf{u}-\frac{1}{24}\mathbf{u}\cdot(\nabla\nabla\mathbf{u})\cdot\mathbf{u}\right),\end{equation}
where ${\bf u} = {\bf u}({\bf z}(t))$.

Writing the error per time step as $e=\Delta^{2}w(t)$, where $w(t)=\left|(\mathbf{u}\cdot\nabla\mathbf{u})\cdot\nabla\mathbf{u}/12-\mathbf{u}\cdot(\nabla\nabla\mathbf{u})\cdot\mathbf{u}/24\right|$, the total error in an
integration is\begin{equation}
E=\int_{0}^{T}edt=\int_{0}^{T}\Delta^{2}w(t)dt.
\label{eq:local-total-error-physical}
\end{equation}
Now if we write the variable time step $\Delta t$ as $\Delta=(dt/d\tau)h$, where $h=\Delta\tau$ is the constant step size in $\tau$,
this error takes the form
\begin{equation}
E=h^{2}\int d\tau\left(\frac{dt}{d\tau}\right)^{3}w(t).
\label{eq:local-global-error}
\end{equation}

We proceed to minimize the total error $E$ by writing $E=h^{2}\int d\tau\mathcal{L}(t,dt/d\tau)$.
Ignoring the constant factor $h^2$, the Lagrangian $\mathcal{L}$ is given by
\begin{equation}
\mathcal{L}(t,dt/d\tau)=\left(dt/d\tau\right)^{3}w(t),
\end{equation}
where $\tau$ is the time, and we wish to find $t(\tau)$ that minimizes
$E$. This is done by forming the Euler-Lagrange equations, or equivalently by
doing a Legendre transform to a Hamiltonian:
\begin{eqnarray}
p=\frac{\partial\mathcal{L}}{\partial(dt/d\tau)}=3\left(\frac{dt}{d\tau}\right)^{2}w(t),\\
\mathcal{H}(t,p)=p\left(\frac{dt}{d\tau}\right)-\mathcal{L}(t,dt/d\tau).
\end{eqnarray}
We obtain $
\mathcal{H}= 2 (p/3)^{3/2}  w(t)^{-1/2}$.
Notice that the Hamiltonian $\mathcal{H}$ is independent of the time
$\tau$ and is therefore a constant of motion. Also, the Hamiltonian
$\mathcal{H}$ is proportional to the Lagrangian $\mathcal{L}=(dt/d\tau)^{3}w(t)=(p/3)^{3/2} w(t)^{-1/2}$,
implying that $\mathcal{L}$ is also a constant.  Since $dt/d\tau = 1/\rho$,
we obtain an expression for the optimized time steps:
\begin{equation}
\rho_{opt}(\mathbf{z}(t))=C w(t)^{1/3}.\label{eq:equidistribution}
\end{equation}
This condition says that the Lagrangian, the integrand in Eq.~(\ref{eq:local-global-error}),
is constant with respect to $\tau$. (But the local error $e$ in
Eq.~(\ref{eq:local-total-error-physical}) is not constant with respect
to $t$.) A condition of this form is
called an \emph{equidistribution principle}\cite{liseikingrid}.

\subsection{Numerical results}
\label{sec:num_opt}
\begin{figure}[htbp]
\begin{center}
\includegraphics[scale=.8]{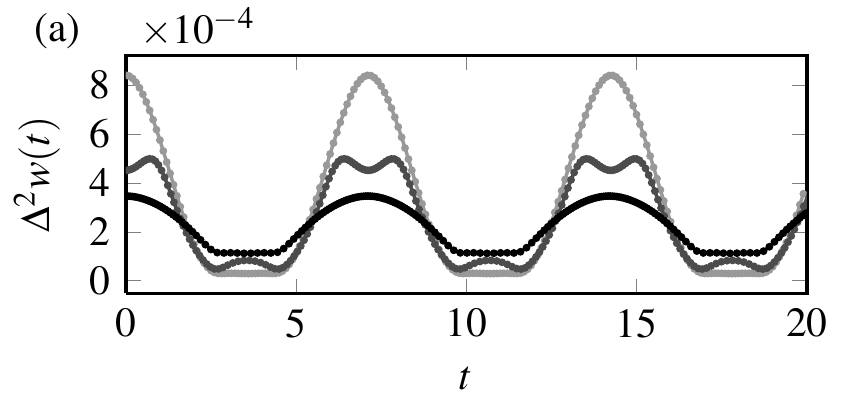}
\includegraphics[scale=.8]{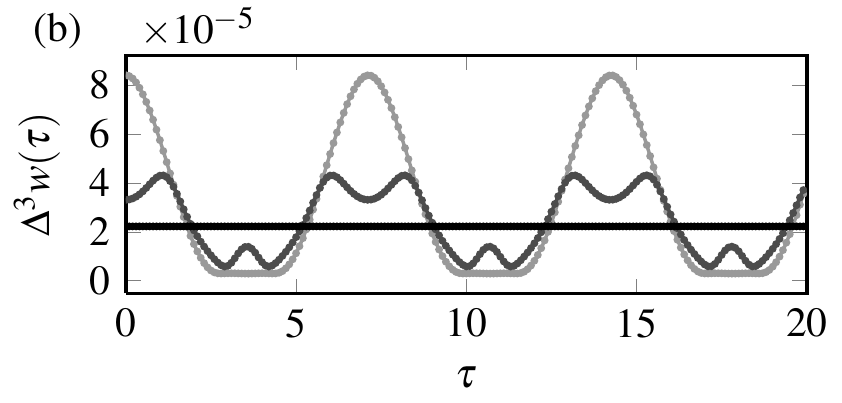}
\caption{\label{fig:errors}
Comparison of the errors for the extended phase space CN method, for three time-step variations, $\Delta=\rm{const.}$ (light gray), $\Delta_{arc}$ (dark gray), and $\Delta_{opt}$ (black).  (a)~The local error as a function of time $t$ [integrand of Eq.~\eqref{eq:local-total-error-physical}], and (b)~the local error as a function of~$\tau$ [$h^3$~times the integrand of Eq.~\eqref{eq:local-global-error}].  Parameters: $h=0.1$, $(q_0,p_0) = (0.4,0)$, $\Delta_0=0.1$, $C_{arc} = 0.41$, $C_{opt}=0.28$
}
\end{center}
\end{figure}

In order to test the optimized time step, we compare numerical results using constant time step [$\Delta=\Delta_0$], the optimized time step 
\begin{eqnarray}
\Delta_{opt} = C_{opt} w(q,p)^{-1/3} \propto 1/\rho_{opt}  ,
\end{eqnarray}
and time steps designed to give equal arc length:
\begin{eqnarray}
\Delta_{arc} =C_{arc} \left[ \left(\frac{\partial H}{\partial p}\right)^2 + \left(\frac{\partial H}{\partial q}\right)^2 \right]^{-1/2}.
\end{eqnarray}
With $\Delta_{arc}$, the magnitude of the phase space velocity $\vec v \equiv (\partial q/\partial\tau, \partial p/\partial \tau)$ is constant.  Numerical results for these three choices of $\Delta$ are shown in Fig.~\ref{fig:errors}, using the CN integrator in extended phase space.  For each of these methods, fixed steps in $\tau$ of size $\Delta \tau = h$ were used.  The constant of proportionality $C$ in $\Delta = C/\rho$ was then chosen so that, after $N$ periods of the orbit, $\tau(N)$ equals $t(N)$.  This is done in order to distinguish the effect of varying the step size as a function of $(q,p)$ from the effect of reducing the overall average step size.  

Two diagnostics of the numerical integrations related to the local error are shown in Fig.~\ref{fig:errors}, for $\Delta=\Delta_0$ (light gray), $\Delta_{arc}$ (dark gray), and $\Delta_{opt}$ (black).  They are the local error $\Delta^2 w(t)$ as a function of time and $\Delta^3 w(\tau)$, proportional to the integrand in Eq.~\eqref{eq:local-global-error}, as a function of $\tau$, respectively.  Variations in step size given by $\Delta_{arc}$ reduce the error compared to the fixed step size integration, at almost all points along the orbit.  Variations given by $\Delta_{opt}$ further reduce the local error in some places, while allowing the local error to increase in others.  As seen in Fig.~\ref{fig:errors}b, this variation has the effect of spreading the local error along the orbit, so that $\Delta^3 w(t)$ is constant as a function of $\tau$, as required by the equidistribution principle.

\begin{figure}[htbp]
\begin{center}
\includegraphics[scale=.9]{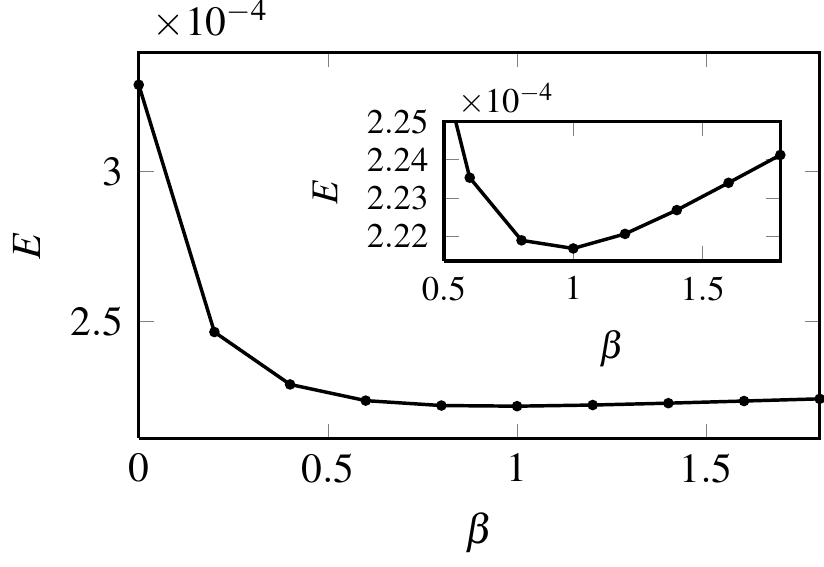}
\caption{\label{fig:opt1}
The global error [Eq.~\eqref{eq:local-global-error}], over the time $T=20$, for the time step variation $\Delta_\beta$ [Eq.~\eqref{eq:opt_scan}].  As seen in the inset, the error is minimized with $E=2.22\times 10^{-4}$ for $\beta=1$, for which $\Delta_\beta=\Delta_{opt}$.  For comparison, using $\Delta=\Delta_{arc}$ gives $E=2.48\times 10^{-4}$, and $\Delta=\Delta_0$ gives $E=3.29\times10^{-4}$.
}
\end{center}
\end{figure}

As a further test to check if $\Delta_{opt}$ really minimizes $E$ [as defined in Eq.~(\ref{eq:local-global-error})], we also show numerical results where $E$ is computed for 
\begin{eqnarray}\label{eq:opt_scan}
\Delta_\beta &=& C_{\beta} \left[ (1-\beta) + \beta w(q,p)^{-1/3} \right],\\
&\propto& \left[ (1-\beta)\Delta_0 + \beta\Delta_{opt}\right],
\end{eqnarray}
where $C_{\beta}$ is chosen as above for each $\beta$, for a range of $\beta$.  When $\beta=0$, the time steps are uniform, and when $\beta=1$ the time steps are given by $\Delta_{opt}$.  The results are shown in Fig.~\ref{fig:opt1}.   As seen in the inset in Fig.~\ref{fig:opt1}, the error is indeed a minimum for $\beta=1$, with $E=2.22\times10^{-4}$.  For comparison, using $\Delta=\Delta_{arc}$ gives $E=2.48\times 10^{-4}$, and $\Delta=\Delta_0={\rm const.}$ gives $E=3.29\times10^{-4}$

\section{Summary}
\label{sec:summary}

In this paper we have studied variable and adaptive time stepping for symplectic integrators, of importance for particle-in-cell codes, for accelerators, for tracing magnetic field lines, and for ray tracing.
We have shown that problems observed in the literature for symplectic integrators
with variable time steps fall into two categories. In the first, with
time step $\Delta=\Delta(t)$, the integrators are still symplectic
but results can exhibit parametric instabilities associated with resonances
between the time step variation and the orbital motion. We have characterized
these instabilities by means of backward error analysis and numerical
integrations. In the second category, the time steps depend on position
in phase space, $\Delta=\Delta(q,p)$, and integrators which are symplectic for
$\Delta={\rm const.}$ are no longer symplectic. 

We have discussed symplectic integrators for $\Delta=\Delta(q,p)$ of two basic
varieties. In both, we start by modifying the time variable $t\rightarrow\tau$
such that $\rho(q,p)dt=d\tau$ and $\rho(q,p)\propto\Delta(q,p)^{-1}$ and we do uniform time stepping in $\tau$.
In the first method, one extends the phase space $(q_{1},...,q_{n},p_{1},...,p_{n})\longrightarrow(q_{0},q_{1},...,q_{n},p_{0},p_{1},...,p_{n})$,
with $q_{0}=t$ and $p_{0}$ its canonical conjugate. In this extended
phase space with $\tau$ taking the place of time and $\Delta\tau={\rm const.}$,
any numerical integrator can be used and we investigate modified leapfrog\cite{finn:054503,fichtl2011},
symmetrized to be of second order, and Crank-Nicolson. In the second
variety of integrator, we start by noticing that the equations to be solved, written as functions of $\tau$, are in Hamiltonian form with non-canonical
variables for one degree of freedom problems.  We then construct a Poisson integrator
(the analog of a symplectic integrator but for non-canonical variables)
in terms of a generating function of mixed variables, which is a generalization
of the mixed variable generating functions for canonical variables\cite{Goldstein:2002sf}.
We also show how to symmetrize this integrator to obtain second order
accuracy, obtaining the non-canonical symmetrized leapfrog integrator.

We have investigated these integrators for a model Hamiltonian
system (the cubic oscillator) and found no parametric
instabilities or growth or decay due to lack of phase space conservation.

We have also investigated adaptive time-stepping, focusing on minimizing
the norm of the error obtained by backward error analysis. We have
applied this optimal time-stepping to the model Hamiltonian system
and found that indeed the error is minimized while the advantages
of symplectic integration are preserved.

\section*{Acknowledgments}

We wish to thank A.~Dragt, P.~Morrison, and K.~Rasmussen for enlightening discussions,
and R.~Pich\'e for sharing a copy of his unpublished report~\cite{piche98}.

This work was funded by the Laboratory Directed Research and Development program (LDRD), under the auspices of the U. S. Department of Energy by Los Alamos National Laboratory, operated by Los Alamos National Security LLC under contract DE-AC52-06NA25396.

\appendix

\section{Variable time steps and perturbation theory}
\label{sec:pert}

Two main time step variations are considered in the literature: $\Delta = \Delta(t)$ and $\Delta = \Delta(q,p)$.  The phase space dependent variation $\Delta(q,p)$ is natural because the errors in an integration method of an autonomous system depend only on phase space position.  Indeed, we found that error analysis (as in Sec.~\ref{sec:h_qp_error_est}) for the integration method gives errors which depend on location in phase space $(q,p)$.

The second kind of variation $\Delta(t)$ comes from a type of perturbation analysis.  The logic is as follows: orbits $(q(t),p(t))$ of a Hamiltonian system are often periodic, and because of this, time step variations $\Delta(q,p)$ on the orbit should also be periodic.  That is, we can substitute $q(t)$ and $p(t)$ to arrive at $\Delta(t)\equiv \Delta(q(t),p(t))$, giving periodic time dependence.

This argument ignores the stability issue: if an orbit is perturbed, the period of $\Delta(q(t),p(t))$ will vary, but this will not occur for a prescribed $\Delta(t)$.

It has been observed\cite{skeel_1993,lee1997variable,Wright1998421,Skeel1998758} that such a $\Delta(t)$ is not a stable method for varying the time step of a symplectic integrator.  Some analysis (see Ref.~\cite{piche98} and Sec.~\ref{sec:h_t}) shows that this does not work because it introduces parametric resonances into the system.  As we have shown in Sec.~\ref{sec:h_t}, the integrator is still symplectic in this case, but the problem is that the modified equations (in the case of a harmonic oscillator) take the form of the Mathieu equation, which is still Hamiltonian but can be unstable.

We here review how such parametric instabilities and secularities arise in perturbation theory, to help us understand the parametric instabilities of Sec.~\ref{sec:h_t_analytic}. 

Secularities show up in perturbation theories in the following manner. Take the system
\begin{equation}
\ddot{x}+x+\epsilon x^{3}=0.\label{eq:x-cubed-system}
\end{equation}
A straightforward perturbation method involves iterating
\begin{equation}
\ddot{x}_{k+1}+x_{k+1}=-\epsilon x_{k}^{3}.
\end{equation}
For $x_{0}=b \cos t$, we obtain
\begin{equation}
\ddot{x}_{1}+x_{1}=-\epsilon b^{3}\left(\frac{1}{4}\cos 3t+\frac{3}{4}\cos t\right).
\end{equation}
The last term on the right is resonant and leads to a secularity,
\begin{equation}
x_{1}=b \cos t-\frac{3\epsilon b^{3}}{8}t \sin t+\cdots
\end{equation}
However, it is clear from Eq.~(\ref{eq:x-cubed-system}) that the
exact solution is bounded: the energy is $H=\dot{x}^{2}/2+x^{2}/2+\epsilon x^{4}/4$.
The $t \sin t$ term represents the frequency shift that occurs
in Eq.~(\ref{eq:x-cubed-system}) because of the term proportional
to $\epsilon$, and is accurate for short time, but clearly wrong
for long time.

For an example of a perturbation approach with a parametric instability,
we consider instead
\begin{equation}
\ddot{x}+x+\epsilon x^{2}=0,
\end{equation}
with the perturbation method
\begin{equation}
\ddot{x}_{k+1}+x_{k+1}=-\epsilon x_{k}x_{k+1}.
\label{eq:pert_2}
\end{equation}
(We choose a different model in order to obtain again results at first order in the perturbation theory.) From $x_{0}=b \cos t$ we obtain 
\begin{equation}
\ddot{x}_{1}+\left(1+\epsilon b \cos t\right)x_{1}=0.
\end{equation}
Using $\tau=t/2$, this is the Mathieu equation $d^{2}x_{1}/d\tau^{2}+\left(a+2q \cos 2\tau\right)x_{1} = 0$,
with $a=4$ and $q=2\epsilon b$, for which a parametric instability
occurs. As in the secular example above, the exponential growth represents
the correct frequency shift for small time, but is wrong for longer
time. Indeed, the energy is $H=\dot{x}^{2}/2+x^{2}/2+\epsilon x^{3}/3$,
and for initial conditions near $x=\dot{x}=0$ the solution cannot
grow indefinitely. 
As we have seen in Sec.~\ref{sec:h_t_analytic}, parametric instabilities for $\Delta t=\Delta(t)$ have a very similar character.

\section{Non-canonical brackets and forms}
\label{sec:app_a}

In this appendix we first describe the conditions that change of variables must satisfy so that the form of a set of equations defined with a non-canonical bracket is preserved.  We then show that the flow of a Hamiltonian system written in non-canonical variables, such as Eq.~\eqref{eq:non-canonical} (one degree of freedom), generates a change of variables satisfying these conditions.

First define the non-canonical bracket
\begin{eqnarray}
[f({\bf z}),g({\bf z})]_z \doteq \frac{1}{\rho({\bf z})} \epsilon_{ij} \frac{\partial f({\bf z})}{\partial z_{i}} \frac{\partial g({\bf z})}{\partial z_{j}}.
\end{eqnarray}
In one degree of freedom, $[\cdot, \cdot]_z$ is indeed a non-canonical bracket, which can be verified by proving the Jacobi identity. In more that one degree of freedom, this does not define a legitimate non-canonical bracket, since it does not satisfy the Jacobi identity.

We now ask what change of variables $\mathbf{Z}=\mathbf{Z}(\mathbf{z})$
preserves the form, i.e., what conditions must this change of variables satisfy so that
\begin{eqnarray}\label{eq:pres_req}
[f({\bf z}),g({\bf z})]_z = [f({\bf Z}),g({\bf Z})]_Z,
\end{eqnarray}
where
\begin{eqnarray}
[f({\bf Z}),g({\bf Z})]_Z \doteq  \frac{1}{\rho({\bf Z})} \epsilon_{mn} \frac{\partial f({\bf Z})}{\partial Z_{m}} \frac{\partial g({\bf Z})}{\partial Z_{n}}.
\end{eqnarray}
Using the definitions of these brackets and the chain rule, Eq.~\eqref{eq:pres_req} becomes
\begin{equation}
\frac{1}{\rho({\bf z})} \epsilon_{ij} 
\frac{\partial Z_m}{\partial z_{i}} 
\frac{\partial f({\bf z})}{\partial Z_m} 
\frac{\partial Z_n}{\partial z_{j}} 
\frac{\partial g({\bf z})}{\partial Z_n} 
= 
\frac{1}{\rho({\bf Z})} \epsilon_{mn} 
\frac{\partial f({\bf Z})}{\partial Z_m} 
\frac{\partial g({\bf Z})}{\partial Z_n} .
\end{equation}
In order for this to be true, we require that
\begin{equation}\label{eq:pres_req2}
\frac{1}{\rho({\bf z})} \epsilon_{ij} 
\frac{\partial Z_m}{\partial z_{i}} 
\frac{\partial Z_n}{\partial z_{j}} 
= 
\frac{1}{\rho({\bf Z})} \epsilon_{mn},
\end{equation}
or, written in matrix form, that
\begin{equation}
MEM^{T}=\frac{\rho(\mathbf{z})}{\rho(\mathbf{Z})}E,\,\,\,\,{\rm where}\,\,\,\, M_{ij}=\frac{\partial Z_{i}}{\partial z_{j}},\,\,\,\,\,\,\, E_{ij}=\epsilon_{ij}.\label{eq:MEM-transpose}
\end{equation}
If this condition is satisfied, then the transformation ${\bf z} \to {\bf Z}$ preserves this specific non-canonical symplectic structure.

The next issue is to show that the time evolution operator for Eq.~\eqref{eq:non-canonical}
generates
a map with this property. 
Since Eq.~\eqref{eq:non-canonical} can be written in terms of the non-canonical bracket as
\begin{eqnarray}
\frac{dz_i}{d\tau} = [z_i, H]_z,
\end{eqnarray}
then, for infinitesimal $h$, if we define $z_{i}=z_{i}(0)$
and $Z_{i}=z_{i}(h)$ we have
\begin{equation}
Z_{i}=z_{i}+h[z_{i},H]_{z}.
\end{equation}
Taking the bracket of $Z_i$ and $Z_j$ therefore gives
\begin{eqnarray}
[Z_{i},Z_{j}]_{z}&=[z_{i},z_{j}]_{z}+h[[z_{i},H]_{z},z_{j}]_{z}+h[z_{i},[z_{j},H]_{z}]_{z} \\
&=[z_{i},z_{j}]_{z}+h[[z_{i},H]_{z},z_{j}]_{z}+h[[H,z_{j}]_{z},z_{i}]_{z} \\
&=[z_{i},z_{j}]_{z}+h[[z_{i},z_{j}]_{z},H]_{z}.
\end{eqnarray}
The last equality follows from the Jacobi identity. This equals
\begin{eqnarray}
[Z_{i},Z_{j}]_{z}&=\epsilon_{ij}\left(\frac{1}{\rho(\mathbf{z})}+h[{1}/{\rho},H]_{z}\right) \\
&=\epsilon_{ij}\left(\frac{1}{\rho(\mathbf{z})}+(Z_{k}-z_{k})\frac{\partial}{\partial z_{k}}\left(\frac{1}{\rho(\mathbf{z})}\right)\right) \\
&=\frac{\epsilon_{ij}}{\rho(\mathbf{Z})}=[Z_{i},Z_{j}]_{Z}.
\end{eqnarray}
Thus, the time evolution operator for an infinitesimal time step $h$
preserves the non-canonical symplectic structure. This fact is the
basis for finding a non-canonical generating function in Sec.~\ref{sec:noncan} that
gives a first order accurate integrator for the non-canonical Hamiltonian
system.

Finally, we express these ideas in terms of differential forms. 
Eq.~\eqref{eq:pres_req2} can be written
\begin{equation}
\rho(\mathbf{Z})\frac{\partial Z_{k}}{\partial z_{i}}\epsilon_{kl}\frac{\partial Z_{l}}{\partial z_{j}}=\rho(\mathbf{z})\epsilon_{ij}.\end{equation}
From this, we obtain the expression 
\begin{equation}
\rho(\mathbf{Z})(\partial Z_{k}/\partial z_{i})\epsilon_{kl}(\partial Z_{l}/\partial z_{j})dz_{i}dz_{j}=\rho(\mathbf{z})\epsilon_{ij}dz_{i}dz_{j},
\end{equation}
or 
\begin{equation}
\rho(\mathbf{Z})\epsilon_{kl}dZ_{k}dZ_{l}=\rho(\mathbf{z})\epsilon_{ij}dz_{i}dz_{j}.
\end{equation}
Since we are considering only 2D phase space, this can be expressed as 
\begin{equation}
\rho(X,Y)dX\wedge dY=\rho(x,y)dx\wedge dy.\label{eq:non-canonical-two-form}
\end{equation}
Eq.~(\ref{eq:non-canonical-two-form}) is the two-form used in Sec.~\ref{sec:noncan}.

{~}

\section*{References}


\end{document}